\documentclass[prd,preprint,tightenlines,showpacs,aps,amsmath,amssymb,nofootinbib]{revtex4} 
\usepackage{graphicx} 
\usepackage{epstopdf}
\DeclareGraphicsExtensions{.pdf,.eps,.png,.jpg,.mps}  
\usepackage{epsfig}

\newcommand{\beq}{\begin{equation}}
\newcommand{\eeq}{\end{equation}}

\begin{document}

\title{Cosmological Constraints on Strongly Coupled Moduli from Cosmic Strings} 

\author{Eray Sabancilar}
\email{eray.sabancilar@tufts.edu}
\affiliation{Institute of Cosmology, Department of Physics and Astronomy,
Tufts University, Medford, MA  02155, USA.}

\begin{abstract}
Cosmological constraints on moduli, whose coupling to matter is stronger than Planck mass suppressed coupling, are derived. In particular, moduli are considered to be produced by oscillating loops of cosmic strings and constraints are obtained from their effects on big bang nucleosynthesis and their contribution to diffuse gamma ray background and dark matter. Large volume and warped Type-IIB flux compactifications are taken as examples where strongly coupled moduli are present. Finally, the constraints on cosmic string tension, modulus mass and modulus coupling constant are obtained and it is shown that the constraints are relaxed significantly when the coupling constant is large enough. In addition, the effects of thermal production of moduli are considered and the corresponding constraints are derived.
\end{abstract}

\maketitle


\section{Introduction}
 
String theory requires the presence of scalar fields called moduli such as complex structure and Kahler moduli which parametrize the volume and the shape of a six dimensional manifold representing the extra dimensions in string theory. There is also the modulus called dilaton whose expectation value determines the strength of the string coupling constant. Moduli are originally massless and their values are presumably fixed by the dynamics of the theory so that in the effective theory they become massive scalar fields which are phenomenologically acceptable. Moduli stabilization is still a challenging problem in string theory, however, in flux compactification scenarios, it is possible to fix moduli by turning on fluxes in the internal manifold \cite{GKP}. The possibility of having a large number of values for different fluxes leads to the picture of string theory landscape where there exist $10^{500}$ different vacua. In this large landscape of vacua, there are attractive models where some of the long standing problems are revisited such as the hierarchy \cite{GKP,large-volume1, large-volume2}, the possibility of having a de Sitter vacuum in string theory \cite{KKLT}, brane inflation as the origin of inflation \cite{Dvali-Tye, KKLMMT} and cosmic superstrings \cite{Sarangi-Tye, Dvali-Vilenkin, Polchinski-review}. 

Moduli can be produced by the oscillating loops of cosmic strings\footnote{Moduli can also be produced thermally if the reheating temperature is high enough. We shall comment on that possibility in section \ref{thermal-moduli}.}. Such moduli can have effects on big bang nucleosynthesis (BBN) and can also contribute to dark matter and diffuse gamma ray background. These effects for the gravitationally coupled moduli have been studied in detail \cite{Damour-Vilenkin-moduli, Babichev-Kachelriess, Peloso-Sorbo}. However, recently, some compactification scenarios have been introduced where moduli couple to matter more strongly than the Planck mass suppressed coupling \cite{Frey-Maharana, Burgess, Conlon-Quevedo}. In this paper, we show that the cosmological constraints on moduli become less severe when moduli couplings are stronger.      

In general, moduli were expected to have Planck mass suppressed couplings \cite{DeWolfe-Giddings}. However, in warped and large volume flux compactification scenarios they may couple to matter more strongly. Refs. \cite{Frey-Maharana,Burgess} argue that the dilaton is localized in the IR region of a throat for a large warping. The dilaton mass is suppressed by the warp factor and coupling to matter is stronger than the Planck mass suppressed coupling. Localization of wavefunctions and stronger couplings to matter are expected for other moduli as well \cite{Frey-Maharana,Burgess}. They also show that there is a smooth interpolation between moderate and large warping cases which mimics the Randall-Sundrum (RS) model \cite{Randall-Sundrum} as the effective theory in the large warping case. 

The Giddings-Kachru-Polchinski model \cite{GKP} was the first string theory realization of producing large hierarchies from pure numbers, i.e., quanta of fluxes. It was argued that the RS model gives an effective description of the warped compactification scenario with a large warp factor where the bulk space is replaced by the UV brane and all the $4$D physics except for gravity is localized on the IR brane located at the bottom of the throat \cite{Brummer}. In the original RS model, the radial modulus is not fixed and left as a free parameter. A mechanism for stabilizing this modulus was proposed by Goldberger and Wise \cite{Goldberger-Wise}, who showed that this modulus has a TeV suppressed coupling rather than Planck mass suppressed and has a TeV scale mass if the hierarchy problem is solved. Brummer et al further showed that the RS model with the radial modulus stabilized by the Goldberger-Wise mechanism is the effective description of the warped compactification scenario \cite{Brummer}. Therefore, there is some evidence for the moduli with strong coupling in the warped Type-IIB flux compactification scenario.  

Another model where a strongly coupled modulus is present is the so called large volume compactification where volume becomes exponentially large\cite{large-volume1, large-volume2}. It was shown in \cite{Conlon-Quevedo} that one of the Kahler moduli can have mass $\sim 10^{6} GeV$ and coupling to matter is suppressed by the string mass scale $m_{s} \sim 10^{11} GeV$ for a particular value of volume which leads to $TeV$ scale SUSY breaking. In this model, there is another Kahler modulus with mass $\sim 1 MeV$ and Planck mass suppressed coupling to matter which suggests the presence of both strongly and weakly coupled moduli together.

In this paper, we derive constraints on both strongly and weakly coupled moduli produced by the oscillating loops of cosmic strings. We obtain upper limits on the abundance of such moduli from diffuse gamma ray background \cite{EGRET}, big bang nucleosynthesis \cite{Kawasaki, Jedamzik, Ellis} and the dark matter density \cite{WMAP}. We also use the lower limit on the scalar field mass from Cavendish-type experiments \cite{Hoyle}. We find the allowed region of the parameter space in terms of string tension $G\mu$, modulus coupling constant $\alpha$ and modulus mass $m$.    

 
\section{Moduli Radiation From Strings}

A modulus $\phi$ couples to matter via trace of its energy momentum tensor
\beq
\mathcal{L}_{int} \sim \frac{\alpha}{m_{p}} \phi T_{\mu}^{\mu},
\eeq
where $\alpha$ is the modulus coupling constant, $m_{p}$ is the Planck mass and $T_{\mu}^{\mu}$ is the trace of the matter energy momentum tensor.

We consider oscillating loops of cosmic strings coupled to a modulus as a periodic source of moduli production. Moduli radiation from a loop of cosmic string occurs with the power \cite{Damour-Vilenkin-moduli}
\beq
P_{m} \sim 30 \alpha^{2} G \mu^{2},
\eeq
when the loop size $L \lesssim 4\pi / m$ where m is the modulus mass. This part of the spectrum corresponds to moduli produced from small oscillating loops and so it is relevant to the early universe. We shall call this part of the spectrum as background moduli. The corresponding average particle emission rate is
\beq\label{Ndot}
\dot{N} \sim 13 \frac{ \alpha^{2} G \mu^{2}}{\omega},
\eeq
where $\omega$ is the energy of a modulus in the rest frame of the loop. The moduli are mainly produced in the fundamental oscillation mode with $\omega = 4 \pi /L$ where $L$ is the size of the loop \cite{Damour-Vilenkin-moduli}. Thus, the particle emission rate can be expressed as
\beq
\dot{N} \sim \frac{13}{4 \pi} \alpha^{2} G \mu^{2}L.
\eeq

When $L >> 4\pi / m$, the main contribution to the radiation spectrum comes from cusps and has a different power spectrum. Such moduli are produced in late epochs and have larger lifetimes due to large boost factors of the cusps. Possible observable effects of such moduli will be worked out in \cite{Berezinsky-Sabancilar-Vilenkin}. Here, we shall only consider the background moduli and their cosmological effects. 


\section{Lifetime and Loop Density}

The rate of decay of a modulus into the standard model (SM) gauge bosons can be estimated as
\beq
\Gamma \sim n_{SM} \left(\frac{\alpha}{m_{p}}\right)^{2} m^{3},
\eeq
where $n_{SM} =12$ is the total number of spin degree of freedom for all SM gauge bosons and $m$ is the modulus mass. The mean lifetime of such a modulus in its rest frame can be estimated as the inverse of the decay rate as
\beq
\tau \sim 8.1 \times 10^{12} \alpha^{-2} m_{GeV}^{-3}\,s,
\eeq
where $m_{GeV} \equiv m/1\, GeV$.

An oscillating loop of cosmic string also produces gravitational radiation with the power \cite{Vilenkin-book}
\beq
P_{g} \sim 50 G \mu^{2}.      
\eeq

The main energy loss mechanism for a loop of cosmic string is the gravitational radiation provided that $P_{g} \gtrsim P_{m}$ which corresponds to the case when $\alpha \lesssim 1$. $\alpha \sim 1$ case was worked out in \cite{Damour-Vilenkin-moduli, Babichev-Kachelriess, Peloso-Sorbo}. However, when moduli are strongly coupled to matter, i.e., $\alpha >> 1$, the dominant energy loss mechanism becomes the moduli radiation and so this leads to significant modifications of the constraints obtained in \cite{Damour-Vilenkin-moduli, Babichev-Kachelriess, Peloso-Sorbo}. 

The lifetime of a loop when the moduli radiation dominates is
\beq\label{tau_L}
\tau_{L} \sim \frac{\mu L}{P_{m}} \sim \frac{L}{30 \alpha^{2} G \mu}.
\eeq

The constraints we shall obtain in the next section depend upon the length of the loops formed from the cosmic string network. There is still no consensus on the evolution of string network and analytical works \cite{Dubath,Vanchurin} and different simulations \cite{Martins,Ringeval,Vanchurin-Olum-Vilenkin,Olum-Vanchurin} yield different answers. However, the biggest recent simulations \cite{Vanchurin-Olum-Vilenkin,Olum-Vanchurin} suggest that a loop formed at cosmic time $t$ has a typical length
\beq
L \sim \beta t,
\eeq
with $\beta \sim 0.1$.

The loops of interest to us are the ones formed in the radiation dominated era whose number density is given by \cite{Vilenkin-book}
\beq\label{rad-loop-density}
n(L,t)\sim \zeta \beta^{1/2} (t L)^{-3/2},
\eeq
where $\zeta \sim 16$, $30 \alpha^{2} G \mu t \lesssim L \lesssim \beta t$.

Loops lose their energy mostly via friction in the friction dominated epoch when $t \lesssim t_{*} \sim t_{p}/(G\mu)^{2}$ \cite{Vilenkin-book} and cannot produce moduli efficiently. Therefore, we consider later times where the loops reach a scaling solution and the main energy loss mechanism is via moduli radiation.  

The particle emission rate (\ref{Ndot}) is valid for the loops of size $L \lesssim 4\pi/m$, which exist only at $t \lesssim t_{m}$. From (\ref{tau_L}) we obtain $t_{m}$ as
\beq\label{t_m}
t_{m} \sim \frac{4\pi}{30} \alpha^{-2} (G \mu)^{-1} m^{-1}.
\eeq

Thus, we shall be interested in moduli produced in the time interval 
\beq\label{time-constraint}
t_{*} \lesssim t \lesssim t_{m}.
\eeq
This implies $t_{m} \gtrsim t_{*}$, which can be expressed as
\beq\label{friction-constraint}
G \mu \gtrsim \frac{30}{4\pi} \alpha^{2} \frac{m}{m_{p}} \sim 2 \times 10^{-19} \alpha^{2} m_{GeV}.
\eeq
We represent this condition on the parameter space plots given in the next section as a dashed line below which moduli are not produced, hence there are no constraints on moduli in that region\footnote{Plasma friction may or may not affect cosmic F- and D-strings depending on whether they interact with ordinary matter or not. However, thermally produced bulk field background, such as moduli, might have a similar effect on cosmic F- and D-strings. If they are not affected by friction, the condition (\ref{friction-constraint}) is removed for cosmic F- and D-strings.}.


\section{Cosmological Constraints on Moduli}\label{constraints}


\subsection{Abundance}

Moduli abundance is $Y(t) = n_{m}(t) / s (t)$ where $n_{m}(t)$ is the moduli number density and $s(t)$ is the entropy density given by 
\beq
s(t) = 0.0725 \mathcal{N}^{1/4} \left(\frac{m_{p}}{t}\right)^{3/2},
\eeq
where $\mathcal{N} \sim 100$ is the total number of spin degrees of freedom in the radiation dominated era at time $t$.

The total number of moduli produced by a single loop until cosmic time $t < \tau_{L}$ can be obtained from (\ref{Ndot}) as  
\beq\label{N}
N \sim \dot{N} t \sim \frac{13}{4\pi} \alpha^{2} G \mu^{2} L t.
\eeq 
By using (\ref{rad-loop-density}) and (\ref{N}), the number density of moduli in the universe produced by the loops of size $L$ can be found as
\beq
n_{m}(t) \sim N n(L, t) \sim \frac{13}{4\pi} \zeta \beta^{1/2} (Lt)^{-1/2} \alpha^{2} G \mu^{2}.
\eeq
Thus, the moduli abundance can be estimated as
\beq
Y \sim 4.5 \zeta \beta^{1/2} L^{-1/2} t m_{p}^{-3/2} \alpha^{2} G \mu^{2},
\eeq
Note that the smallest loops of size $L \sim 30 \alpha^{2} G \mu t$ dominate the abundance. It can also be seen from (\ref{time-constraint}) that $t = t_{m}$ gives the most dominant contribution to the abundance. Using these facts, we obtain
\beq\label{abundance}
Y \sim 2.7 (G\mu) \left(\frac{m_{p}}{m}\right)^{1/2} \sim 9.4 \times 10^{9} (G \mu) m_{GeV}^{-1/2}.
\eeq
Note that the dependence on $G \mu$ in equation (\ref{abundance}) is different from that found in \cite{Damour-Vilenkin-moduli} since there, it was assumed that $\beta \sim G \mu$. Although $\beta \sim 0.1$ seems to lead to more stringent constraints on the string tension $G \mu$, we shall see that the constraints are relaxed when the coupling constant $\alpha$ becomes large enough.  


\subsection{Cosmological Constraints on Strongly Coupled Moduli}

Short distance measurements of Newton's Law of gravity in Cavendish type experiments give a lower bound on the modulus mass as $m> 10^{-3} eV$, i.e., $m_{GeV} > 10^{-12}$ \cite{Hoyle}.
\begin{figure}\label{figure1}
  \begin{center}
    \begin{tabular}{cc}
\epsfysize=6cm \epsfbox{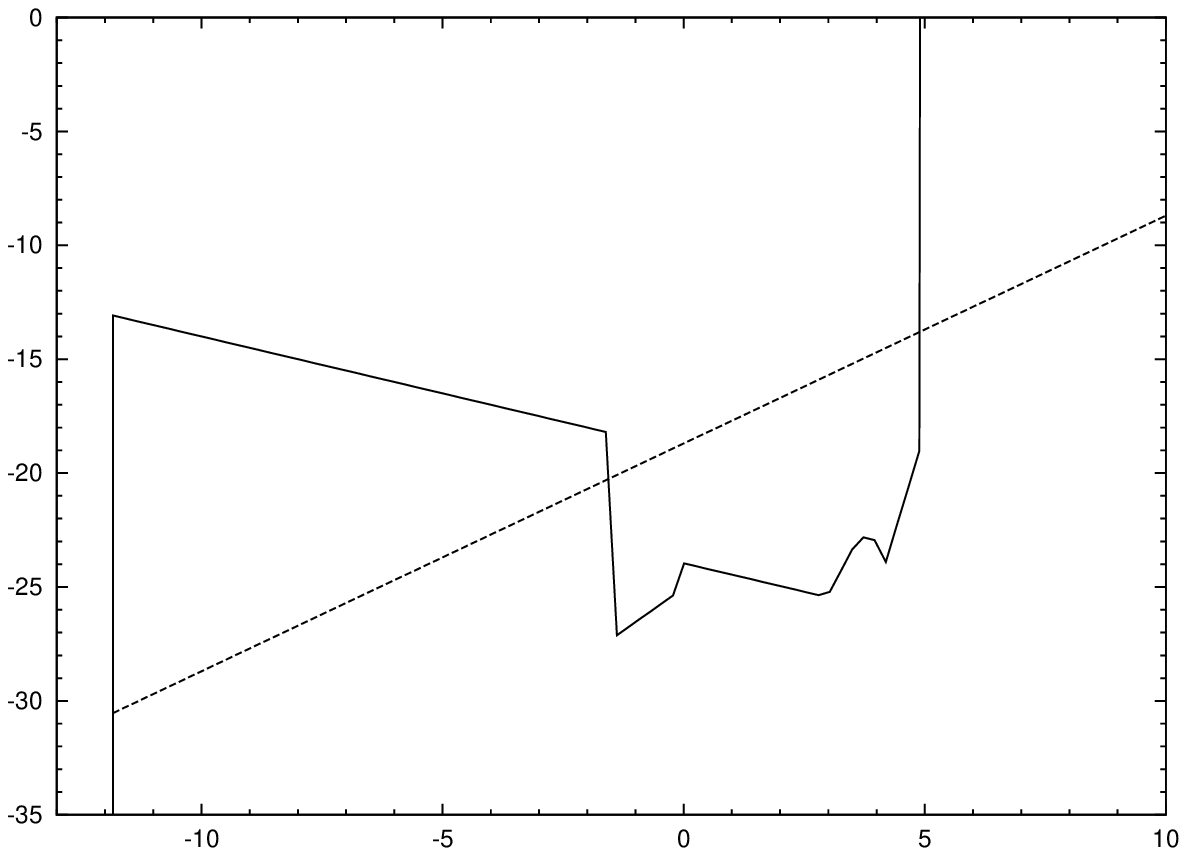}&
\put(-260,85){\scriptsize{log$\,G \mu$}}
\put(-125,-10){\scriptsize{log$\,m_{GeV}$}}
\put(-215,150){\scriptsize{$\alpha=1$}}
\epsfysize=6cm \epsfbox{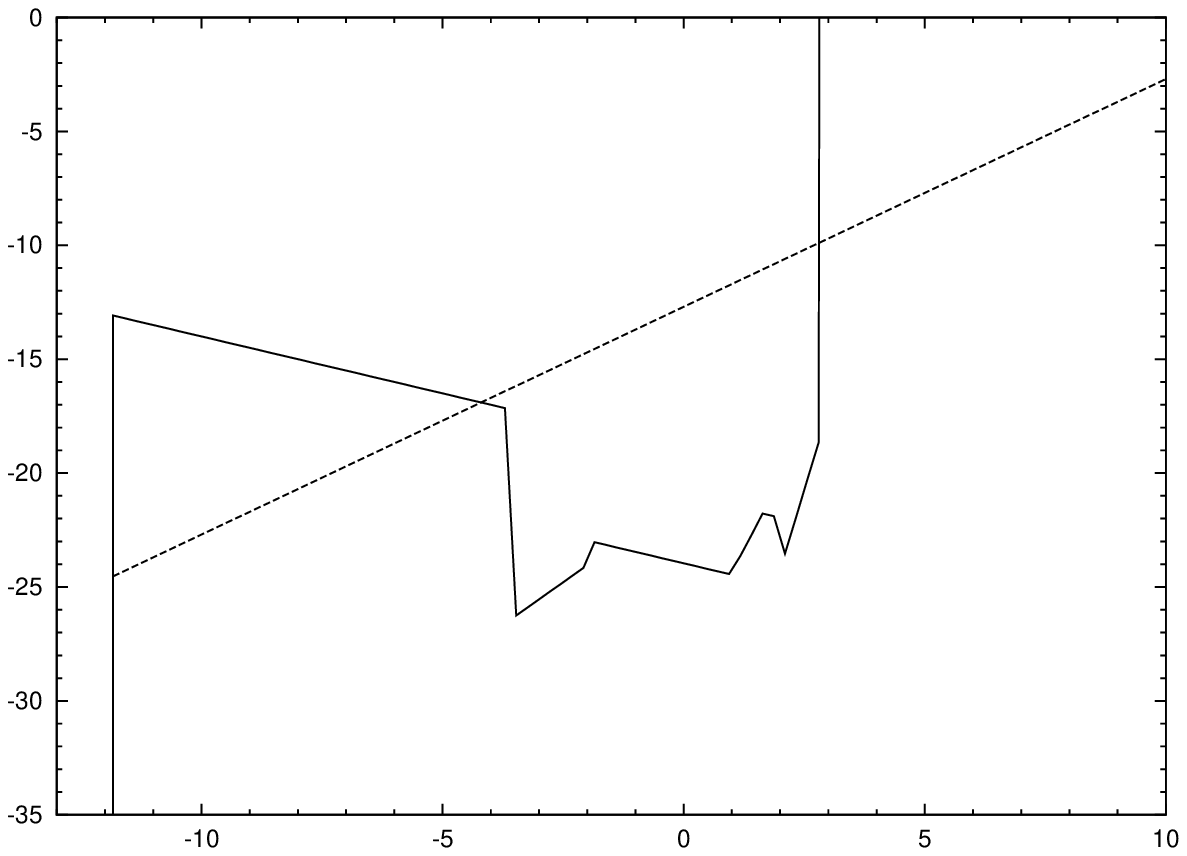}
\put(-249,85){\scriptsize{log$\,G \mu$}}
\put(-125,-10){\scriptsize{log$\,m_{GeV}$}}
\put(-215,150){\scriptsize{$\alpha=10^{3}$}}\\
\epsfysize=6cm \epsfbox{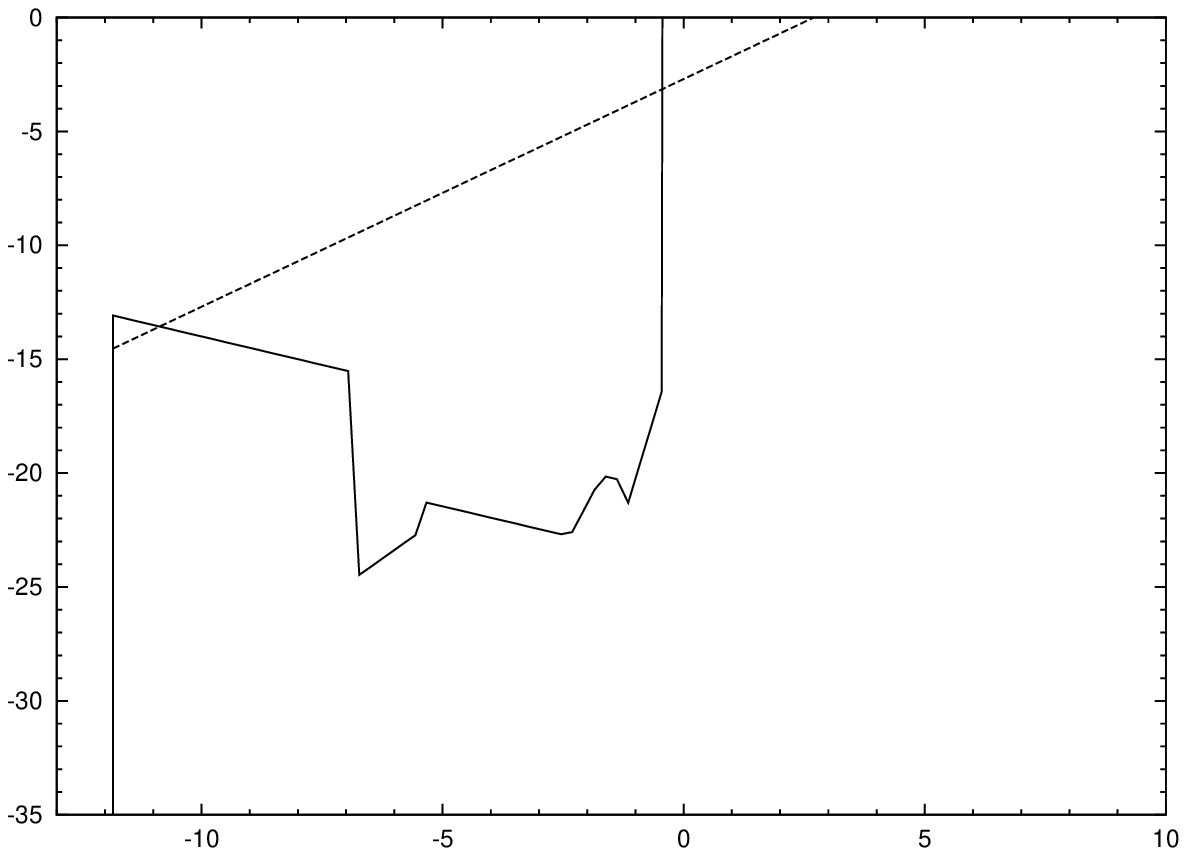}
\put(-260,85){\scriptsize{log$\,G \mu$}}
\put(-125,-10){\scriptsize{log$\,m_{GeV}$}}
\put(-215,150){\scriptsize{$\alpha=10^{8}$}}\\
\end{tabular}
\caption{log$\,G\mu$ vs log$\,m_{GeV}$ for strongly coupled moduli. The region above the solid line is forbidden by the cosmological constraints and the region below the dashed line is free of the constraints for the loops affected by plasma friction since such moduli are never produced because of friction domination. Note that if F- and D- strings do not interact with ordinary matter like solitonic cosmic strings do, the friction domination does not apply. Hence, one should ignore the dashed line in that case.}
\end{center}
\end{figure}

If moduli are long-lived, i.e., $\tau \gtrsim t_{0} \sim 4.3 \times 10^{17} s$, they contribute to the dark matter in the universe. Thus, we have the upper bound $\Omega_{m} h^{2} < 0.13$ \cite{WMAP} or in terms of abundance $Y < 9.6 \times 10^{-10} m_{GeV}^{-1}$. 

If moduli are long-lived, they also contribute to the diffuse gamma ray background \cite{Berezinsky}. When $\tau \gtrsim t_{0}$, the energy density of moduli that decayed into photons until the present time can be estimated as
\beq
\rho_{m} \sim Y s(t_{0}) m \frac{t_{0}}{\tau} \sim 2.2 \times 10^{17} Y m_{GeV}^{4} \alpha^{2}\,\, eV\, cm^{-3},
\eeq
where $t_{0}/\tau$ is the fraction of the decayed moduli and $s(t_{0}) = s(t_{eq}) (t_{eq}/t_{0})^{2} \sim 2.9 \times 10^{-38} GeV^{3}$. According to EGRET data, an approximate upper bound on the diffuse gamma ray density for the photons of energy $>1\,MeV$ is $\rho_{\gamma} \sim 2 \times 10^{-6}\,\, eV\, cm^{-3}$ \cite{EGRET}. Using this upper bound, we can estimate the limit on the abundance from the constraint $\rho_{m} \lesssim \rho_{\gamma}$ as $Y \lesssim 9.1 \times 10^{-24} \alpha^{-2} m_{GeV}^{-4}$.

When $t_{dec} \sim 10^{13}\, s \lesssim \tau \lesssim t_{0}$, the most stringent constraint comes from the diffuse gamma ray background \cite{Berezinsky}. Assuming all the moduli decay by the time $\tau$, the energy density can be estimated as 
\beq
\rho_{m} \sim Y s(\tau) m \sim 1.1 \times 10^{22} Y \alpha^{4} m_{GeV}^{7}\,\, eV\, cm^{-3},
\eeq
where $s(\tau) = s(t_{eq}) (t_{eq}/\tau)^{2} \sim 8.2 \times 10^{-29} \alpha^{4} m^{6}\, GeV^{3}$. Redshifting the photon energy density to time $t=\tau$, we find 
\beq
\rho_{\gamma}(\tau) \sim \rho_{\gamma} \left(\frac{t_{0}}{\tau}\right)^{8/3} \sim 1 \times 10^{7} \alpha^{16/3} m^{8} \, eV\, cm^{-3},  
\eeq
which gives the upper bound on the moduli abundance as $Y \lesssim 9.1 \times 10^{-16}\,\alpha^{4/3} m_{GeV}$.

If the modulus lifetime is shorter than $t_{dec}$, they can have effects on BBN \cite{Kawasaki, Jedamzik, Ellis}. When such moduli decay electromagnetically, they dissolve the light elements created during nucleosynthesis. Besides, moduli-gluon coupling leads to hadron production which can change the primordial light element abundances. To obtain upper limits on the moduli abundance, we made a piecewise power law approximation to the results of \cite{Kawasaki, Jedamzik, Ellis} and summarized them in table \ref{table1}, where $\tau_{s} \equiv \tau/sec$.
\begin{table}[ht]
\begin{center}
\vspace{3mm} 
\begin{tabular}{|c|c|c|cl}
\hline
$\tau_{s}$ & $Y$  \\
\hline 
$10^{4} \lesssim \tau_{s} \lesssim 10^{13}$ & $10^{-14} m_{GeV}^{-1}$\\
\hline 
$10^{2} \lesssim \tau_{s} \lesssim 10^{4}$ & $10^{-8} \tau_{s}^{-3/2} m_{GeV}^{-1}$\\
\hline
$10 \lesssim \tau_{s} \lesssim 10^{2}$ & $10^{-11} m_{GeV}^{-1}$\\
\hline
$10^{-2} \lesssim \tau_{s} \lesssim 10$ & $10^{-11} \tau_{s}^{-5/2} m_{GeV}^{-1}$\\
\hline
\end{tabular}
\caption{BBN constraints on the strongly coupled moduli abundance. This table shows the approximate upper bounds on the strongly coupled moduli abundance as a function of modulus lifetime and modulus mass.}
\label{table1}
\end{center}
\end{table}

Using the bounds obtained from Cavendish-type experiments, diffuse gamma ray background, BBN and dark matter constraints, we obtain the limits on string tension $G\mu$, modulus mass $m$ and modulus coupling constant $\alpha$. Using all the constraints, we obtained Fig.1 for the parameter space of $G \mu$ vs $m_{GeV}$ for several values of $\alpha$. The analytic form of the constraints in all parameter ranges is given in Table \ref{table2} of the Appendix. As it can be seen from Fig.1, the constraints become weaker as $\alpha$ increases. The condition (\ref{friction-constraint}) moves up on the parameter space which leads to the region free of constraints below the dashed line where no moduli are produced from cosmic strings.
\begin{figure}\label{figure2}
  \begin{center}
    \begin{tabular}{cc}
\epsfysize=6cm \epsfbox{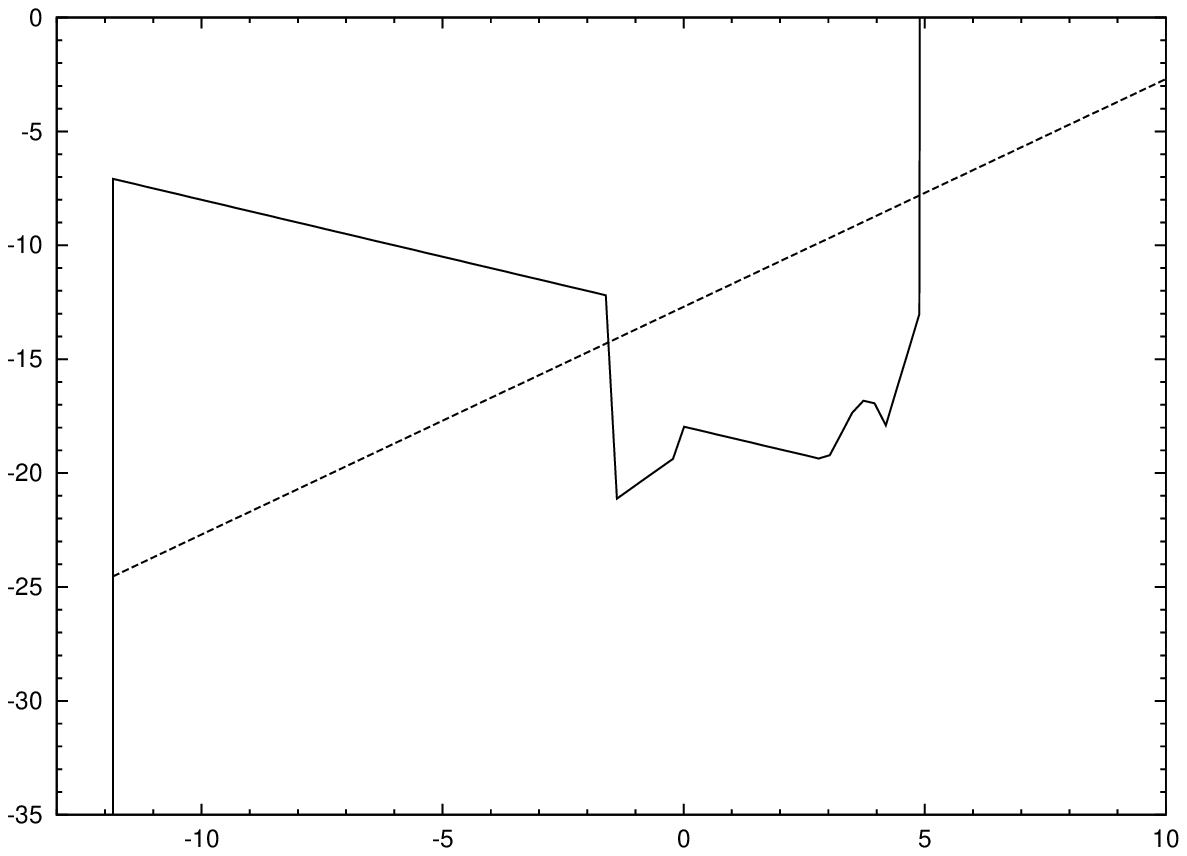}&
\put(-260,85){\scriptsize{log$\,G \mu$}}
\put(-125,-10){\scriptsize{log$\,m_{GeV}$}}
\put(-215,150){\scriptsize{$\alpha=10^{3}$}}
\put(-215,143){\scriptsize{$\alpha_{W}=1$}}
\epsfysize=6cm \epsfbox{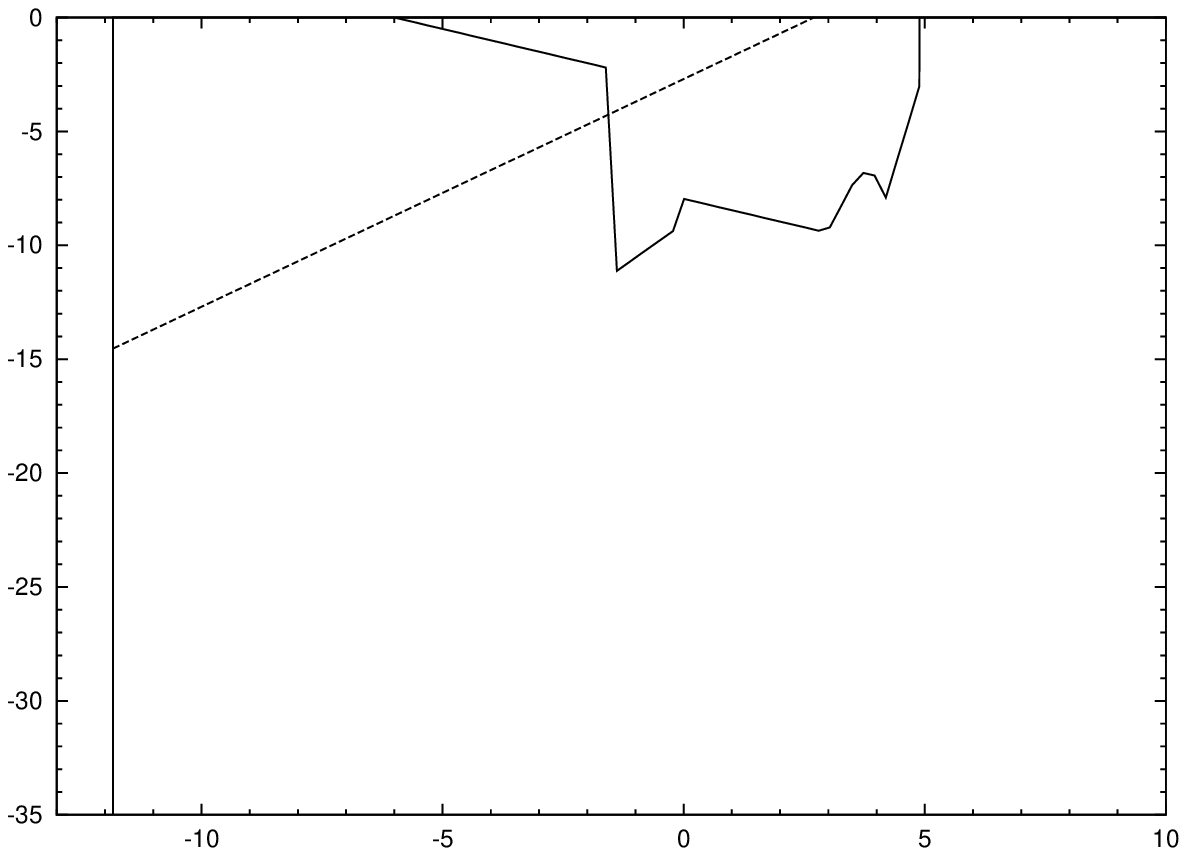}
\put(-249,85){\scriptsize{log$\,G \mu$}}
\put(-125,-10){\scriptsize{log$\,m_{GeV}$}}
\put(-208,150){\scriptsize{$\alpha=10^{8}$}}
\put(-208,143){\scriptsize{$\alpha_{W}=1$}}\\
\epsfysize=6cm \epsfbox{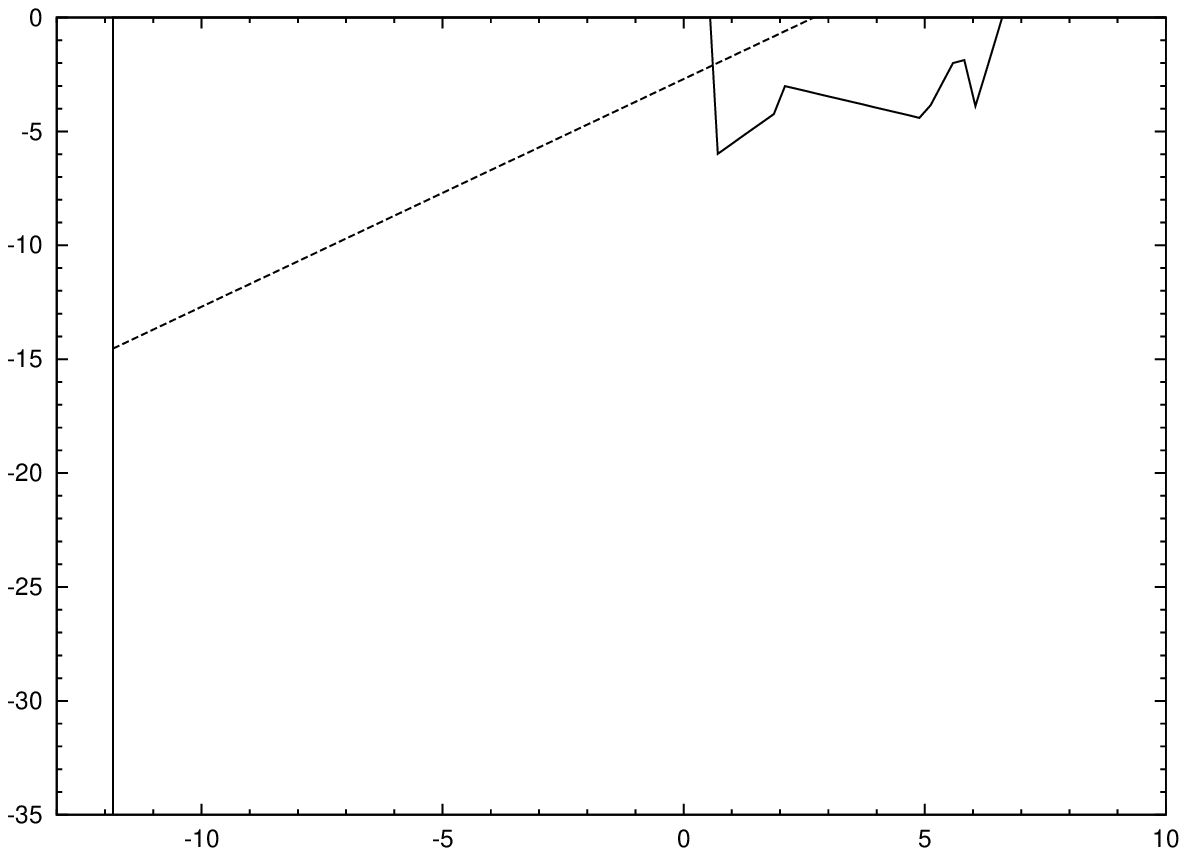}
\put(-260,85){\scriptsize{log$\,G \mu$}}
\put(-125,-10){\scriptsize{log$\,m_{GeV}$}}
\put(-208,150){\scriptsize{$\alpha=10^{8}$}}
\put(-208,143){\scriptsize{$\alpha_{W}=10^{-3}$}}\\
\end{tabular}
\caption{log$\,G\mu$ vs log$\,m_{GeV}$ for weakly coupled moduli when $m_{weak} \gtrsim m_{strong}$. The region above the solid line is forbidden by the cosmological constraints and the region below the dashed line is free of the constraints for the loops affected by plasma friction since such moduli are never produced because of friction domination. Note that if F- and D- strings do not interact with ordinary matter like solitonic cosmic strings do, then the friction domination does not apply. Hence, one should ignore the dashed line in that case.}
\end{center}
\end{figure}


\subsection{Cosmological Constraints on Weakly Coupled Moduli}

In the previous section, we analyzed the cosmological constraints on strongly coupled moduli. In this section, we shall assume that there are at least one strongly coupled modulus and one weakly coupled modulus (coupling suppressed by at least Planck mass) with coupling constants $\alpha>>1$ and $\alpha_{W} \lesssim 1$ respectively. We shall estimate the cosmological constraints on weakly coupled moduli similar to the previous section. Note that the dominant energy loss mechanism for the loops is still via strongly coupled moduli radiation, thus the loop lifetime is given by (\ref{tau_L}) and the minimum size of the loops is $L_{min} \sim 30 \alpha^{2} G \mu t$. On the other hand, the modulus lifetime depends upon its coupling constant to matter and given by
\beq
\tau_{W} \sim 8.1 \times 10^{12} \alpha_{W}^{-2} m_{GeV}^{-3}\,s.
\eeq
The abundance of weakly coupled moduli can be calculated as
\beq
Y_{W} \sim 9.4 \times 10^{9} \alpha^{-2} \alpha_{W}^{2} (G \mu) m_{GeV}^{-1/2}.
\eeq
which is valid for $m_{weak} \gtrsim m_{strong}$ where $m_{weak}$ and $m_{strong}$ are the masses of the strongly and the weakly coupled moduli respectively. However, if $m_{weak} < m_{strong}$, although the strongly coupled moduli production terminates at $t_{m_{strong}}(\alpha)$ given by equation (\ref{t_m}), weakly coupled moduli are still produced and the process terminates at $t_{m_{weak}}(\alpha=1)$. At this point, gravitational radiation starts dominating and the abundance becomes  $Y_{W} \sim 9.4 \times 10^{9} \alpha_{W}^{2} (G \mu) m_{GeV}^{-1/2}$. Therefore, the constraints are the same as given on Fig.1 for $\alpha =1$ case when $m_{weak}<m_{strong}$ assuming $\alpha_{W} \sim 1$.

In the opposite regime, when $m_{weak}>m_{strong}$, by using all the constraints, we show the parameter space in Fig.2 for various values of $\alpha_{W}$ and $\alpha$. Once again, the analytic form of the constraints is given in table \ref{table3} in the Appendix. As it can be seen from Fig.2, the constraints become less important as $\alpha$ increases and $\alpha_{W}$ decreases since the abundance is suppressed by $\alpha^{-2} \alpha_{W}^{2}$. Besides, the condition (\ref{friction-constraint}) becomes stronger for larger $\alpha$ and there is a larger region in the parameter space below the dashed line free of the constraints. 


\section{Thermally Produced Moduli Background}\label{thermal-moduli}

So far we have discussed the production of moduli from cosmic strings. Moduli can also be produced thermally if the reheating temperature is high enough. The photon-modulus interaction can be written as
\beq
\mathcal{L}_{int} \sim \frac{\alpha}{m_{p}} \phi F^{\mu \nu} F_{\mu \nu}.
\eeq

Moduli should be in thermal equilibrium with photons for the thermal production. The lowest order process which contributes to the interaction $\gamma \gamma \to \phi \phi$ is second order and the cross section can be estimated as
\beq
\sigma \sim \left(\frac{\alpha^{2}}{m_{p}^{2}}\right)^{2} E^{2},
\eeq
where $E \sim T$ is the energy of photons at temperature $T$. 

For the thermal production of moduli to occur, the pathlength of moduli should be less than the Hubble time, i.e., 
\beq\label{thermal-production}
\frac{1}{\sigma n_{\gamma}} \lesssim t,
\eeq
where $t$ is the cosmic time and $n_{\gamma}$ is the photon density at temperature $T$. Using $n_{\gamma} \sim T^{3}$ and $t\sim m_{p}/T^{2}$ in (\ref{thermal-production}), we obtain
\beq\label{reheating-temperature}
T \gtrsim \alpha^{-4/3} m_{p}.
\eeq

For instance, when $\alpha \sim 10^{9}$, (\ref{reheating-temperature}) implies $T_{rh} \gtrsim 10^{7} GeV$. Since reheating temperature is also model dependent, strongly coupled moduli may or may not be produced thermally. On the other hand, weakly coupled moduli cannot be produced since $T_{rh} \gtrsim m_{p}$ is required for $\alpha \lesssim 1$. 

Assuming strongly coupled moduli are produced thermally and dominate the universe, we can estimate the temperature after their decay. The decay rate of moduli is
\beq
\Gamma \sim \left(\frac{\alpha}{m_{p}}\right)^{2} m^{3},
\eeq
and when $\Gamma \sim H \sim T^{2}/m_{p}$, moduli will decay and reheat the universe to temperature $T$. Using that, we obtain
\beq
T \sim \alpha \left(\frac{m}{m_{p}}\right)^{1/2} m.
\eeq

The weakest constraint one can consider is that $T$ should be at least at the nucleosynthesis temperature $\sim 1 MeV$. Using $T \gtrsim 1 MeV$, we obtain the constraint
\beq\label{reheating-constraint}
\alpha \gtrsim 10^{6} m_{GeV}^{-3/2}.
\eeq
  

\section{Conclusions}

We consider oscillating loops of cosmic strings as periodic sources of moduli production. When $\alpha \lesssim 1$, gravitational radiation is the dominant energy loss mechanism for the loops. The constraints for this case is identical to $\alpha = 1$ case for the strongly coupled moduli as we have shown on Fig. 1. Note that our results for $\alpha \sim 1$ are more stringent than that of \cite{Damour-Vilenkin-moduli, Babichev-Kachelriess, Peloso-Sorbo}. This is mainly because of the fact that they assume $\beta \sim G\mu$ in their calculations whereas we use $\beta \sim 0.1$ from the recent simulations \cite{Vanchurin-Olum-Vilenkin,Olum-Vanchurin}.

When a modulus is strongly coupled to matter, i.e., $\alpha \gtrsim 1$, the dominant energy loss mechanism for the loops becomes the moduli radiation. Hence, loop lifetimes depend upon $\alpha$. Besides, modulus lifetime shortens as $\alpha$ increased if the modulus mass is kept constant. These two effects make the cosmological constraints we obtained for the strongly coupled moduli less severe. Basically, for moduli to have effects on BBN, and to contribute to dark matter and diffuse gamma ray background, their lifetime should be long enough. This means that smaller mass values of the strongly coupled moduli have effects on cosmology as we can see from Fig.1. 

In addition, loops cannot radiate moduli effectively in the friction dominated epoch since they lose their energy mostly in friction. The condition for friction domination (\ref{friction-constraint}) becomes stronger when $\alpha$ is larger. Therefore, more region of the parameter space is allowed as $\alpha$ is increased since friction domination does not let moduli to be produced by cosmic strings in that region. This may not be the case for F- and D- strings since they may or may not interact with ordinary matter depending on where they are located in the bulk. However, if there is a thermally produced moduli background, a similar effect might occur to F- and D- strings which needs further investigation.          

We consider warped and large volume compactifications as the two examples where at least one strongly coupled modulus is present. In the warped compactification scenario, there is some evidence for moduli localization in long throat regions which leads to stronger coupling to matter and smaller moduli masses \cite{Frey-Maharana, Burgess}. Besides, there is a smooth interpolation between large and moderate warping. This suggests that warped compactification with a long throat can be effectively described by the RS model with its radion stabilized by Goldberger-Wise mechanism \cite{Brummer}. As it was argued some time ago by Goldberger and Wise that, RS radion has TeV suppressed coupling to matter \cite{Goldberger-Wise}. Interpolating our results for this particular case, we see that there is almost no cosmological constraint on the RS radions produced by the cosmic strings.

As a second example, we consider large volume compactification where a strongly coupled modulus is present. One of the Kahler moduli in this scenario has a mass $m \sim 10^{6} GeV$ and a string mass scale suppressed coupling where $m_{s} \sim 10^{11} GeV$ \cite{Conlon-Quevedo}. In our notation, this means that $\alpha \sim 10^{8}$, hence a strongly coupled modulus. As we can see from Fig.1, there is almost no cosmological constraint for this modulus.

In  the large volume compactification scenario, there is another Kahler modulus with Planck mass suppressed coupling to matter, i.e., $\alpha \sim 1$. This suggests the possibility of having at least one strongly coupled and a weakly coupled modulus together. We also calculated the constraints on weakly coupled moduli and the results are shown in Fig.1 with $\alpha =1$ for $m_{weak}<m_{strong}$ and in Fig.2 for $m_{weak} \gtrsim m_{strong}$. In particular, the constraints on the weakly coupled Kahler modulus of the large volume scenario is given in the first plot of Fig.1 ($\alpha =1$) since $m_{weak} \sim 1 MeV < m_{strong} \sim 10^{6} GeV$. Note that in this model, maximum string tension can only be $G\mu \sim 10^{-16}$. If we take $\alpha_{W} \sim 1$ and $m_{weak} \sim 1 MeV$, there are not many constraints on this weakly coupled modulus. On the other hand, as it can be seen from Fig.2, constraints are quite weak for $\alpha =10^{8}$, $\alpha_{W} = 1$ case. If $\alpha_{W}<<1$, then there is almost no constraint on weakly coupled moduli with $m_{weak} \gtrsim m_{strong}$ since their abundance is suppressed by $\alpha^{-2} \alpha_{W}^{2}$. 

We also consider the possibility of producing moduli thermally. We found that if the universe has ever reached temperature of order $T \sim \alpha^{-4/3} m_{p}$, then moduli can be produced thermally. If the hierarchy problem is solved with warped geometry, the RS radion couples to matter with $\alpha \sim 10^{15}$ and (\ref{reheating-constraint}) implies $m \gtrsim 10^{-6} GeV$ for the RS radion mass. Since it is expected that $m \sim 1\,TeV$ \cite{Goldberger-Wise}, the RS radion is free of both the cosmological constraints from cosmic strings and from the thermally produced radion background constraint. For the strongly coupled Kahler modulus in the large volume scenario, the constraint (\ref{reheating-constraint}) implies $m \gtrsim 10^{-2} GeV$. Since $m \sim 10^{6} GeV$ in this model \cite{Conlon-Quevedo}, it is also free of the thermally produced moduli background constraint. Finally, we also found that weakly coupled moduli cannot be produced thermally since $T \gtrsim \ m_{p}$ would be required for this to happen.

In this work, we assumed that the reconnection probability of strings is $p=1$ which is true for the ordinary cosmic strings. However, for cosmic F- and D-strings $p<1$ which leads to an enhancement of the string density in the universe\cite{Dvali-Vilenkin,Polchinski-review}. Therefore, the constraints are expected to be a little bit stronger for $p<1$ case which is not a significant effect \cite{Babichev-Kachelriess}. 

The main conclusion of this work is that when there is at least one type of strongly coupled modulus, both the cosmological and the thermally produced moduli background constraints on strongly and weakly coupled moduli become less severe and for sufficiently large values of $\alpha$, there is almost no constraint on moduli. 


\section{Acknowledgments}

I would like to thank Alexander Vilenkin for many useful comments and for carefully reading the manuscript. This work was supported in part by the National Science Foundation under grant 0855447.

\appendix
\section{Cosmological Constraints on Weakly and Strongly Coupled Moduli}
The analytical forms of the cosmological constraints on string tension $G \mu$ for weakly and strongly coupled moduli are given in Table \ref{table2} and Table \ref{table3} respectively.
\\
\\
\\
\\
\\
\\
\\
\\
\\
\\
\begin{table}[ht]
\begin{center}
\vspace{3mm} 
\begin{tabular}{|c|c|c|cl}
\hline
$m_{GeV}$ & $G \mu$  \\
\hline 
$10^{-12} \lesssim m_{GeV} \lesssim 3\times 10^{-2} \alpha^{-2/3}$ & $1.0 \times 10^{-19} m_{GeV}^{-1/2}$\\
\hline 
$10^{-12} \lesssim m_{GeV} \lesssim 3\times 10^{-2} \alpha^{-2/3}$ & $9.7 \times 10^{-34} \alpha^{-2} m_{GeV}^{-7/2}$\\
\hline
$3 \times10^{-2} \alpha^{-2/3} \lesssim m_{GeV} \lesssim 9.3\times 10^{-1} \alpha^{-2/3}$ & $9.1 \times 10^{-26} \alpha^{4/3} m_{GeV}^{3/2}$\\
\hline
$9.3 \times 10^{-1} \alpha^{-2/3} \lesssim m_{GeV} \lesssim 9.3\times 10^{2} \alpha^{-2/3}$ & $1.1 \times 10^{-24} m_{GeV}^{-1/2}$\\
\hline
$9.3 \times 10^{2} \alpha^{-2/3} \lesssim m_{GeV} \lesssim 4.3\times 10^{3} \alpha^{-2/3}$ & $4.6 \times 10^{-38} \alpha^{3} m_{GeV}^{4}$\\
\hline
$4.3 \times 10^{3} \alpha^{-2/3} \lesssim m_{GeV} \lesssim 9.3\times 10^{3} \alpha^{-2/3}$ & $1.1 \times 10^{-21} m_{GeV}^{-1/2}$\\
\hline
$9.3 \times 10^{3} \alpha^{-2/3} \lesssim m_{GeV} \lesssim 9.3 \times 10^{4} \alpha^{-2/3}$ & $5.7 \times 10^{-54} \alpha^{5} m_{GeV}^{7}$\\
\hline
\end{tabular}
\caption{Constraints on the string tension $G \mu$. This table shows the upper bounds we obtain from Cavendish type experiments, diffuse gamma ray background, BBN and dark matter density constraints on $G \mu$ for strongly coupled moduli as a function of modulus mass $m$ and modulus coupling constant $\alpha$.}
\label{table2}
\end{center}
\end{table}
\begin{table}[ht]
\begin{center}
\vspace{3mm} 
\begin{tabular}{|c|c|c|cl}
\hline
$m_{GeV}$ & $G \mu$  \\
\hline 
$10^{-12} \lesssim m_{GeV} \lesssim 3\times 10^{-2} \alpha_{W}^{-2/3}$ & $1.0 \times 10^{-19} \alpha^{2} \alpha_{W}^{-2} m_{GeV}^{-1/2}$\\
\hline 
$10^{-12} \lesssim m_{GeV} \lesssim 3\times 10^{-2} \alpha^{-2/3}_{W}$ & $9.7 \times 10^{-34} \alpha^{2} \alpha_{W}^{-4} m_{GeV}^{-7/2}$\\
\hline
$3 \times10^{-2} \alpha_{W}^{-2/3} \lesssim m_{GeV} \lesssim 9.3\times 10^{-1} \alpha_{W}^{-2/3}$ & $9.1 \times 10^{-26} \alpha^{2} \alpha_{W}^{-2/3} m_{GeV}^{3/2}$\\
\hline
$9.3 \times 10^{-1} \alpha_{W}^{-2/3} \lesssim m_{GeV} \lesssim 9.3\times 10^{2} \alpha_{W}^{-2/3}$ & $1.1 \times 10^{-24} \alpha^{2} \alpha_{W}^{-2} m_{GeV}^{-1/2}$\\
\hline
$9.3 \times 10^{2} \alpha_{W}^{-2/3} \lesssim m_{GeV} \lesssim 4.3\times 10^{3} \alpha_{W}^{-2/3}$ & $4.6 \times 10^{-38} \alpha^{2} \alpha_{W} m_{GeV}^{4}$\\
\hline
$4.3 \times 10^{3} \alpha_{W}^{-2/3} \lesssim m_{GeV} \lesssim 9.3\times 10^{3}  \alpha_{W}^{-2/3}$ & $1.1 \times 10^{-21} \alpha^{2} \alpha_{W}^{-2} m_{GeV}^{-1/2}$\\
\hline
$9.3 \times 10^{3} \alpha_{W}^{-2/3} \lesssim m_{GeV} \lesssim 9.3 \times 10^{4} \alpha_{W}^{-2/3}$ & $5.7 \times 10^{-54}\alpha^{2} \alpha_{W}^{3} m_{GeV}^{7}$\\
\hline
\end{tabular}
\caption{Constraints on the string tension $G \mu$. This table shows the upper bounds we obtain from Cavendish-type experiments, diffuse gamma ray background, BBN and dark matter density constraints on $G \mu$ for weakly coupled moduli when $m_{weak} \gtrsim m_{strong}$ as a function of modulus mass $m$ and moduli coupling constants $\alpha$ and $\alpha_{W}$. When $m_{weak}<m_{strong}$, one should set $\alpha=1$ in the above table.}
\label{table3}
\end{center}
\end{table}



\begin{thebibliography}{99}

\bibitem{GKP}
S. B. Giddings, S. Kachru and J. Polchinski, ``{\it Hierarchies from fluxes in string compactifications}", Phys. Rev. {\bf D 66}, 106006 (2002).

\bibitem{large-volume1}
V. Balasubramanian, P. Berglund, J. P. Conlon and F. Quevedo ``{\it Systematics of Moduli Stabilisation in Calabi-Yau Flux Compactifications} ", JHEP {\bf 03}, 007 (2005). 

\bibitem{large-volume2}
J. P. Conlon, F. Quevedo and Kerim Suruliz, ``{\it Large-volume Flux Compactifications: Moduli Spectrum and D3/D7 Soft Supersymmetry Breaking} ", JHEP {\bf 08}, 007 (2005). 

\bibitem{KKLT}
S. Kachru, R. Kallosh, A. Linde and S. P. Trivedi, ``{\it De Sitter Vacua in String Theory} ", Phys. Rev. {\bf D 68}, 046005 (2003).

\bibitem{KKLMMT}
S. Kachru, R. Kallosh, A. Linde, J. M. Maldacena, L. P. McAllister and S. P. Trivedi, ``{\it Towards Inflation in String Theory} ", JCAP {\bf 10}, 013 (2003). 

\bibitem{Dvali-Tye}
G. R. Dvali and S. H. H. Tye, ``{\it Brane Inflation} ", Phys. Lett. {\bf B 450}, 72 (1999).

\bibitem{Sarangi-Tye}
S. Sarangi and S. H. H. Tye, ``{\it Cosmic String Production towards the end of Brane Inflation}", Phys. Lett. {\bf B 536}, 185 (2002).

\bibitem{Polchinski-review}
J. Polchinski, ``{\it Introduction to Cosmic F- and D-strings} ", hep-th/0412244.

\bibitem{Dvali-Vilenkin}
G. Dvali and A. Vilenkin, ``{\it Formation and Evolution of Cosmic D Strings}", JCAP {\bf 03}, 010 (2004).

\bibitem{Damour-Vilenkin-moduli}
T. Damour and A. Vilenkin, ``{\it Cosmic Strings and The String Dilaton} ", Phys. Rev. Lett. {\bf78}, 2288 (1997).

\bibitem{Babichev-Kachelriess}
E. Babichev and M. Kachelriess, "{\it Constraining Cosmic Superstrings with Dilaton Emission}", Phys. Lett. {\bf B 614}, 1 (2005).

\bibitem{Peloso-Sorbo}
M. Peloso and L. Sorbo, "{\it Moduli from Cosmic Strings}", Nucl. Phys. {\bf B 649}, 88 (2003).

\bibitem{Frey-Maharana}
A. R. Frey and A. Maharana, ``{\it Warped Spectroscopy: Localization of Frozen Bulk Modes} ", JHEP {\bf 08}, 021 (2006).

\bibitem{Burgess}
C. P. Burgess et al., ``{\it  Warped Supersymmetry Breaking} ", JHEP {\bf 04}, 053 (2008).

\bibitem{Conlon-Quevedo}
J. P. Conlon and F. Quevedo, ``{\it Astrophysical and Cosmological Implications of Large Volume String Compactifications} ", JCAP {\bf 08}, 019 (2007).

\bibitem{DeWolfe-Giddings}
O. DeWolfe and S. B. Giddings, ``{\it Scales and hierarchies in warped compactifications and brane worlds} ", Phys. Rev. {\bf D 67}, 066008 (2003).

\bibitem{Randall-Sundrum}
L. Randall and R. Sundrum, ``{\it A Large Mass Hierarchy from a Small Extra Dimension} ", Phys. Rev. Lett. {\bf 83}, 3370 (1999).

\bibitem{Goldberger-Wise}
W. D. Goldberger and M. B. Wise, ``{\it Phenomenology of a Stabilized Modulus} ", Phys. Lett. {\bf B 475}, 275 (2000).

\bibitem{Brummer}
F. Brummer, A. Hebecker and E. Trincherini, ``{\it The Throat as a Randall-Sundrum Model with Goldberger-Wise Stabilization} ", Nucl. Phys. {\bf B 738}, 283 (2006).

\bibitem{EGRET}
P. Sreekumar et al. (EGRET Collaboration), ``{\it EGRET Observations of the Extragalactic Gamma Ray Emission} Ó, Astrophys. J. {\bf 494}, 523 (1998). 

\bibitem{Kawasaki}
M. Kawasaki, K. Kohri and T. Moroi, ``{\it Hadronic Decay of Late-Decaying Particles and Big-Bang Nucleosynthesis} ", Phys. Lett. {\bf B 625}, 7 (2005), astro-ph/0402490

\bibitem{Jedamzik}
K. Jedamzik, ``{\it Big Bang Nucleosynthesis Constraints on Hadronically and Electromagnetically Decaying Relic Neutral Particles} ", Phys. Rev. {\bf D 74}, 103509 (2006).

\bibitem{Ellis}
J. Ellis, K. A. Olive and E. Vangioni, ``{\it Effects of Unstable Particles on Light-element Abundances: Lithium versus Deuterium and He-3} ", Phys. Lett. {\bf B 619}, 30 (2005).

\bibitem{WMAP}
E. Komatsu et al. (WMAP Collaboration),``{\it  Five-Year Wilkinson Microwave Anisotropy Probe (WMAP) Observations: Cosmological Interpretation}", Astrophys. J. Suppl. {\bf 180}, 330 (2009).

\bibitem{Hoyle}
C. D. Hoyle, et al., ``{\it Sub-milimeter Tests of the Gravitational Inverse-square Law} ", Phys. Rev. {\bf D 70}, 042004 (2004).

\bibitem{Berezinsky-Sabancilar-Vilenkin}
V. Berezinsky, E. Sabancilar and A. Vilenkin, {\it in preparation}.

\bibitem{Vilenkin-book}
E.P.S. Shellard and A. Vilenkin, ``{\it Cosmic Strings and Other Topological Defects} ", Cambridge University Press, Cambridge, England (1994).
\bibitem{Dubath}
F. Dubath, J. Polchinski and J. V. Rocha, ``{\it Cosmic String Loops, Large and Small} ", Phys. Rev. {\bf D 77}, 123528 (2008).

\bibitem{Vanchurin}
V. Vanchurin, ``{\it Cosmic String Loops: Large and Small, but not Tiny} ", Phys. Rev. {\bf D 77}, 063532 (2008).

\bibitem{Martins}
C. J. A. Martins and E. P. S. Shellard, ``{\it Fractal Properties and Small-scale Structure of Cosmic String Networks} ", Phys. Rev. {\bf D 73}, 043515 (2006).

\bibitem{Ringeval}
C. Ringeval, M. Sakellariadou and F. Bouchet, ``{\it Cosmological Evolution of Cosmic String Loops} ", JCAP {\bf 02}, 023 (2007).

\bibitem{Vanchurin-Olum-Vilenkin}
V. Vanchurin, K. D. Olum and A. Vilenkin, ``{\it Scaling of Cosmic String Loops} ", Phys. Rev. {\bf D 74}, 063527 (2006).

\bibitem{Olum-Vanchurin}
K. D. Olum and V. Vanchurin, ``{\it Cosmic String Loops in the Expanding Universe} ", Phys. Rev. {\bf D 75}, 063521 (2007).

\bibitem{Berezinsky}
V. S. Berezinsky, ``{\it Neutrino Astronomy and Massive Long-lived Particles From the Big Bang}", Nucl. Phys. {\bf B 380}, 478 (1992).

\end{thebibliography}
\end{document}